\documentclass[%
 reprint,
superscriptaddress,
 amsmath,amssymb,
 aps,
 prl,
]{revtex4-1}

\usepackage{graphicx}% Include figure files
\usepackage{dcolumn}% Align table columns on decimal point
\usepackage{bm}% bold math
\usepackage{diagbox}
\usepackage{setspace}
\usepackage{color}
\usepackage{ulem}
\begin{document}

%\preprint{APS/123-QED}

%\title{New type of topological superconductors with local magnetic symmetries}% Force line breaks with \\

%\title{ Classification of gapped superconducting wire with local magnetic symmetries }

%\title{ New types of topological superconductors in magnetic groups }

\title{New types of topological superconductors under local magnetic symmetries }

\author{Jinyu Zou}
\affiliation{Wuhan National High Magnetic Field Center $\&$ School of Physics, Huazhong University of Science and Technology, Wuhan 430074, China}
\author{Qing Xie}
\affiliation{Wuhan National High Magnetic Field Center $\&$ School of Physics, Huazhong University of Science and Technology, Wuhan 430074, China}
\author{Zhida Song}
\affiliation{Department of Physics, Princeton University, Princeton, New Jersey 08544, USA}
\author{Gang Xu}
 \email{gangxu@hust.edu.cn}
\affiliation{Wuhan National High Magnetic Field Center $\&$ School of Physics, Huazhong University of Science and Technology, Wuhan 430074, China}

\date{\today}% It is always \today, today,

\begin{abstract}

We classify gapped topological superconducting (TSC) phases of one-dimensional quantum wires with local magnetic symmetries (LMSs),
in which the time-reversal symmetry $\mathcal{T}$ is broken but its combinations with certain crystalline symmetry such as $M_x \mathcal{T}$, $C_{2z} \mathcal{T}$, $C_{4z}\mathcal{T}$, and $C_{6z}\mathcal{T}$ are preserved.
Our results demonstrate that an equivalent BDI class TSC can be realized in the $M_x \mathcal{T}$ or $C_{2z} \mathcal{T}$ superconducting wire,
which is characterized by a chiral $Z^c$ invariant.
More interestingly, we also find two types of totally new TSC phases in the $C_{4z}\mathcal{T}$, and $C_{6z}\mathcal{T}$ superconducing wires, which are beyond the known AZ class, and are characterized by a helical $Z^h$ invariant and $Z^h\oplus Z^c$ invariants, respectively. In the $Z^h$ TSC phase, $Z$-pairs of MZMs are protected at each end. In the $C_{6z}\mathcal{T}$ case, the MZMs can be either chiral or helical, and even helical-chiral coexisting.
The minimal models preserving $C_{4z}\mathcal{T}$ or $C_{6z}\mathcal{T}$ symmetry are presented to illustrate their novel TSC properties and MZMs.
\end{abstract}

\maketitle

%\tableofcontents

\textit{Introduction.}---Topological superconductors (TSCs) are new kind of topological quantum states,
which are fully superconducting gapped in the bulk but support gapless excitations
called Majorana zero modes (MZMs) at the boundaries~\cite{Kitaev2001,Qi2011,Sato2011,Ando2015,Sato2017}.
Analogues of the famous Majorana fermions~\cite{Majorana}, MZMs are their own antiparticles,
and are proposed as the qubits of topological quantum computation
because of their nonlocal correlation and
non-Abelian statistic nature~\cite{Ivanov2001,Kitaev2003,Nayak2008,sato2016}.
Hence, searching for TSC materials with MZMs is now an important topic
in condensed matter physics, and a series of schemes have been proposed in the last decade,
including the proximity effect on the surface of
topological insulators (TIs)~\cite{Fu2008,Qi2010,Linder2010,Wang2012,Xu2014,Wang2015}
and the recently predicted intrinsic superconducting
topological materials~\cite{Fu2010,Hosur2011,Tanaka2012,Fu2014,Hosur2014,
Wangzhijun2015,Xu2016,Zhang2018,Wang2018}.

To identify whether a superconductor is topologically nontrivial, we should first ascertain what topological classification it belongs. The topological classification can
be highly enriched by symmetries
including time-reversal symmetry $\mathcal{T}$,
particle-hole symmetry $\mathcal{P}$, and especially
the crystalline symmetries~\cite{Schnyder2008,Ryu2010,Morimoto2013,
Benalcazar2014,Chiu2016,Cornfeld2019,
Zhang2013,Chiu2013,Shiozaki2014,Fang2017}.
The topology for non-interacting Hamiltonians of the 10 Altland-Zirnbauer (AZ) classes
with or without $\mathcal{T}$ and $\mathcal{P}$ has been well classified~\cite{Schnyder2008,Ryu2010}.
Particularly, the Bogoliubov-de Gennes (BdG)
Hamiltonians of the one-dimensional (1D) superconductors,
with $\mathcal{T}$ breaking or preserving, belong to the D and DIII classes respectively.
In both cases we only have the $Z_2$ classification.
Additional to these local symmetries,
crystalline symmetries are considered for each AZ classes to generalize
the topological classification~\cite{Morimoto2013,Benalcazar2014,Chiu2016,Cornfeld2019},
and the topological crystalline superconductors
protected by mirror reflection symmetry~\cite{Zhang2013,Chiu2013}
or rotational symmetries ~\cite{Shiozaki2014,Fang2017}
have been proposed.
Furthermore,
Refs.~\cite{Mizushima2012,Mizushima2013,Fang2014,Shiozaki2014}
have discussed the TSC phase protected by the
magnetic symmetries $M_x\mathcal{T}$ and $C_{2z}\mathcal{T}$.
Nevertheless, the topological classification of
superconductors with general magnetic symmetries is
still an open question, and the corresponding theoretical
analysis is necessary for understanding and searching for new magnetic TSC materials and MZMs.

In this paper, we focus on the topological phases of gapped superconducting wires with local magnetic symmetries (LMSs), in which $\mathcal{T}$ is broken, but its combinations with certain crystalline symmetries, including $M_x \mathcal{T}$, $C_{2z} \mathcal{T}$, $C_{4z}\mathcal{T}$, and $C_{6z}\mathcal{T}$ are preserved.
Our analysis show that, with $M_x\mathcal{T}$ or $C_{2z}\mathcal{T}$ symmetry, an effective BDI class TSC can be realized, which is characterized by a chiral $Z^c$ topological invariant and protects an integer number of MZMs at each end.
Remarkably, two totally new TSC phases are discussed in the superconducting wire with $C_{4z}\mathcal{T}$ or $C_{6z}\mathcal{T}$ symmetry.
In the $C_{4z}\mathcal{T}$ case, the BdG Hamiltonian is characterized by a helical $Z^h$ invariant, which can protects $Z$-pairs of MZMs at each end. The BdG Hamiltonian with $C_{6z}\mathcal{T}$ symmetry possesses the $Z^h\oplus Z^c$ invariants, which means the helical and chiral MZMs can coexist in a single wire system.
The minimal models with the LMS $C_{4z}\mathcal{T}$ and $C_{6z}\mathcal{T}$ are presented separately, in which the TSC with helical MZMs, and the TSC with helical-chiral coexisting MZMs are discussed. Our results may facilitate the ongoing search for novel TSCs.

\begin{figure}
  \centering
  \includegraphics[width=1\columnwidth]{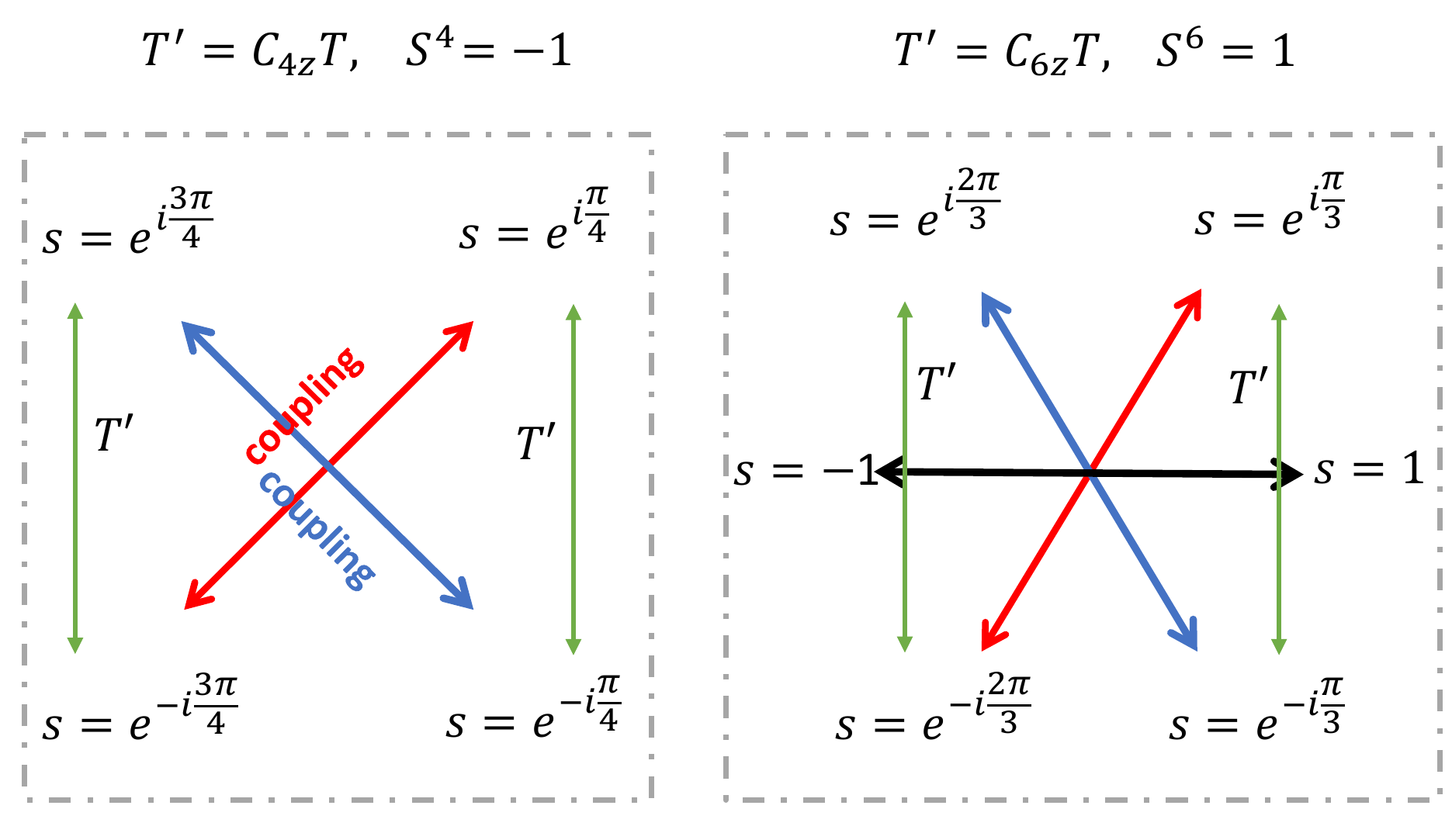}\\
  \caption{The eigenvalues of $\mathcal{S}$ and their transformations in the $C_{4z}\mathcal{T}$ and $C_{6z}\mathcal{T}$ cases. Complex conjugating partners $s$ and $s^*$ are related by the LMSs and always coexist. A perturbation term can be introduced to couple the chiral states with opposite eigenvalues, as illustrated by the red, blue and black double-head arrows.}\label{Fig1}
\end{figure}

\textit{Topological classification of gapped SC wire.}---We
first introduce the LMSs for a magnetic
superconducting wire along $z$ direction.
Among the 1D space groups (the so-called rod group)~\cite{Kopsky2002},
the local symmetry operators include the mirror reflection
$M_x$ and the $n$-fold rotation $C_{nz}$ with $n=2,3,4,6$.
Combined with $\mathcal{T}$,
we obtain four types of LMSs
$\mathcal{T}^\prime =$ $M_x\mathcal{T}$, $C_{2z}\mathcal{T}$,
$C_{4z}\mathcal{T}$ and $C_{6z}\mathcal{T}$, as tabulated in Table~\ref{Table1}.
We consider a 1D BdG Hamiltonian preserving $\mathcal{T}^\prime$.
Notice that the operation of $\mathcal{T}^\prime$ does not change the position of electrons.
Hence it acts on the BdG Hamiltonian as a usual time-reversal operator
\begin{equation}\label{eq:eq1}
  \mathcal{T}^\prime H_{\text{BdG}}(k) \mathcal{T}^{\prime -1} = H_{\text{BdG}}(-k).
\end{equation}
Here, LMS $\mathcal{T}^\prime$
takes the form $\mathcal{T}^\prime=U\mathcal{K}$
with $\mathcal{K}$ being the complex conjugate operator
and $U$ being an unitary matrix determined by the spatial operation and spin flipping.
We take the convention $[\mathcal{T},\mathcal{P}] = 0$
and set $\mathcal{P} = \tau_x \mathcal{K}$,
where the Pauli matrix $\tau_x$ acts on the particle-hole degree of freedom.
Combining $\mathcal{T}^\prime$ and $\mathcal{P}$ leads to
a chiral symmetry $\mathcal{S}=\mathcal{T}^\prime \mathcal{P}$.
$\mathcal{P}$ and $\mathcal{S}$ act on BdG Hamiltonian as follows
\begin{align}
  \mathcal{P}H_{\text{BdG}}(k)\mathcal{P}^{-1} &= -H_{\text{BdG}}(-k), \label{eq:eq2}\\
  \mathcal{S}H_{\text{BdG}}(k)\mathcal{S}^{-1} &= -H_{\text{BdG}}(k). \label{eq:eq3}
\end{align}

The chiral symmetry $\mathcal{S}$
has a series of eigenvalue pairs $\pm s_1, \pm s_2,\cdots$ and
it can take a block-diagonal form as
$\mathcal{S} =\text{diag}[\mathcal{S}_{\pm{s_1}}, \mathcal{S}_{\pm{s_2}}, \cdots]$,
where the subscript $\pm{s_1}$ denotes the direct sum of
eigenvector spaces $|s_1\rangle$ and $|-s_1\rangle$.
The anticommute relation Eq.~(\ref{eq:eq3})
means that $H_{\text{BdG}}(k)$ can be block diagonalized according to the eigenvalues of
$\mathcal{S}^2$.
In other words, $H_{\text{BdG}}(k)$  can adopt the form
$H_{\text{BdG}}(k) = \text{diag} [ H_{s_1^2}, H_{s_2^2}, \cdots]$.
Hence,
%the problem of classifying the superconducting wire Hamiltonian $H_{\text{BdG}}(k)$
%is recast into classifying the Hamiltonian of each block
%and examing the compatibility of
%different block Hamiltonians.
the topological classification of the whole Hamiltonian is
decomposed into examining the topology of
each block and their compatibility.
For each block Hamiltonian $H_{s^2}$,
its topology
is equivalent to either the BDI or the AIII classes,
depending on the chiral symmetry eigenvalue $s$.
To be specific,
when $s$ is a real number,
$H_{s^2}$ is invariant under
$\mathcal{T}'$ or $\mathcal{P}$,
which means it belongs to the BDI class and possesses a $Z$ invariant expressed as $v=N_{s}-N_{-s}$, where $N_{\pm s}$ are the numbers of MZMs with chiral symmetry eigenvalue $\pm s$, respectively.
Additionally,
when $s$ is a complex number, $H_{s^2}$
is transformed into $H_{s^{\ast2}}$ under $\mathcal{T}^\prime$ or $\mathcal{P}$.
Hence, the $H_{s^2}$ ($H_{s^{\ast2}}$) belongs to the AIII
class that is characterized by a $Z$ invariant
$v = N_{s} - N_{-s}$ ($= N_{s^\ast} - N_{-s^\ast}$),
which equals to the number of MZM pairs on each wire end.

%Be aware of that
%the MZMs from different blocks may couple to each other and be destroyed.
%Therefore,
%we need to consider the coupling between the blocks and study its impact on the MZMs.
We next consider the compatibility between the MZMs with different $\mathcal{S}$ eigenvalues.
To do this, we introduce a coupling
term $m|s_1\rangle \langle s_2|$,
which satisfies
$  \mathcal{S} m|s_1\rangle \langle s_2 | \mathcal{S}^{-1} = - m|s_1\rangle \langle s_2 |
 = ms_1 s_2^\ast |s_1\rangle \langle s_2 | $.
%\begin{equation}\label{eq:eq4}
%   \mathcal{S} m|s_1\rangle \langle s_2 | \mathcal{S}^{-1} = - m|s_1\rangle \langle s_2 |.
%\end{equation}
Here $m$ is a perturbation parameter,
and $|s_1\rangle$, $|s_2\rangle$ are the
eigenstates of $\mathcal{S}$.
%\begin{equation}
%  \mathcal{S} m|s_1\rangle \langle s_2 | \mathcal{S}^{-1} = ms_1 s_2^\ast |s_1\rangle \langle s_2 |.
%\end{equation}
Then we see that $m$ can be nonzero only when $s_1 s_2^\ast = -1$,
which means that
MZMs within one block
having chiral eigenvalues $s$ and $-s$
can couple to each other and be eliminated.
However, MZMs from different blocks
are non-interfering due to the protection of $\mathcal{S}$.
Therefore,
the topological classification of the whole BdG Hamiltonian
is determined by the summation of the topology for each block.
We summarize the topological classification of 1D gapped superconductors in
Table~\ref{Table1} and analyse each case in the following.

\newcommand{\tabincell}[2]{\begin{tabular}{@{}#1@{}}#2\end{tabular}} % line break in table
\begin{table}
  \centering
  \caption{The topological classification of the 1D gapped superconducting systems with the LMSs
$M_x\mathcal{T}$, $C_{2z}\mathcal{T}$, $C_{4z}\mathcal{T}$ and $C_{6z}\mathcal{T}$ respectively.
$2\times$AIII form a helical $Z^h$ classification.
}\label{Table1}
  \begin{spacing}{1.2}
  \begin{tabular}{c|cccc}
     \hline
     \diagbox{}{$\mathcal{T}^\prime$} & \tabincell{c}{$M_x \mathcal{T}$\\($n=2$)} & \tabincell{c}{$C_{2z}\mathcal{T}$\\($n=2$)} & \tabincell{c}{$C_{4z}\mathcal{T}$\\($n=4$)} & \tabincell{c}{$C_{6z}\mathcal{T}$\\($n=6$)} \\
     \hline
      $\mathcal{T}^{\prime n}$ & 1 & 1 & $-$1 & 1\\
      $\mathcal{P}^2$ & 1 & 1  & 1 & 1 \\
      $\mathcal{S}^n$ & 1 & 1 & $-$1 & 1 \\
     \hline
     \  Invariant\ \  & \tabincell{c}{\quad $Z^c$ \quad \\ (BDI) } & \tabincell{c}{\quad $Z^c$ \quad \\ (BDI) } & \tabincell{c}{\quad $Z^h$ \quad \\ (2$\times$AIII)} & \tabincell{c}{\quad $Z^h$$\oplus$ $Z^c$ \quad \\ (2$\times$AIII $\oplus$ BDI)}\\
     \hline
   \end{tabular}
   \end{spacing}
\end{table}

(i)~\textit{$M_x\mathcal{T}$ and $C_{2z}\mathcal{T}$ cases}.
---These two cases are equivalent to the BDI class with
$\mathcal{T}^{\prime 2}=1$ and $\mathcal{S}^2=1$.
The chiral topological invariant $v=N_1-N_{-1}\in Z$ is given by the winding number~\cite{Schnyder2008,Sato2017}
\begin{equation}\label{eq:eq5}
  v=\frac{1}{2\pi}\int dk \text{Tr}[W^\dagger(k)\partial_k W(k)] =\frac{1}{2\pi} \int dk \partial_k \theta(k).
\end{equation}
Here $W(k)$ is a unitary matrix that diagonalizes the BdG Hamiltonian
and $\theta (k) $ is the phase angle of $\text{Det}[W(k)]$.
The identity $\text{Tr}[\text{ln}(W)] = \text{ln}(\text{Det}[W])$ is used to derive the above equation.
These results agree well with previous conclusions in references~\cite{Mizushima2012,Mizushima2013,Fang2014,Shiozaki2014,Tewari2012,Samokhin2017}.

(ii)~\textit{$C_{4z}\mathcal{T}$ case}.
--- The chiral symmetry satisfies $\mathcal{S}^4=-1$
and has eigenvalues $\pm e^{\pm i\pi/4}$ (see Fig.~\ref{Fig1}).
One can conclude that the topological invariants are given by $v = N_{e^{i\pi/4}} - N_{-e^{i\pi/4}}$
(or $N_{e^{-i\pi/4}} - N_{-e^{-i\pi/4}})$.
The TSC phase is hence characterized by the helical topological invariant $v \in Z$, which means that the MZMs always appear in Kramers pairs. This is obviously different from the chiral $Z$ invariant in the BDI class, in which the MZMs can arise one by one as $Z$ increasing.
To distinguish the chiral $Z$ and helical $Z$ invariants, we will use $Z^c$ and $Z^h$ in the following.
The $Z^h$ TSC phase of the $C_{4z}\mathcal{T}$-preserving wire can be understood from the following perspective.
The BdG Hamiltonian can be block diagonalized into two sectors according to the eigenvalues $\pm i$ of $C_{2z}=(C_{4z}\mathcal{T})^2$ as
$H_i(k) \oplus  H_{-i}(k)$.
%\begin{equation}\label{eq:eq6}
%    H_{\text{BdG}}^{C_{4z}\mathcal{T}}(k) = \left(
%                 \begin{array}{cc}
%                   H_{i}(k) &  \\
%                    & H_{-i}(k) \\
%                 \end{array}
%               \right).
%\end{equation}
Both $C_{4z}\mathcal{T}$ and $\mathcal{P}$ can map these two sectors to each other.
However, their combination, i.e the chiral symmetry $\mathcal{S}$,
keeps each sector invariant.
As a consequence, each sector belongs to the AIII class, whose $Z^c$ topological invariant can be calculated by exploiting Eq.~(\ref{eq:eq5}).
Yielding to the $C_{4z}\mathcal{T}$ symmetry, the $Z^c$ invariants of two sectors must be equal, which finally gives a $Z^h$ invariant for the whole BdG Hamiltonian.

(iii)~\textit{$C_{6z}\mathcal{T}$ case}.--- We have $\mathcal{T}^{\prime 6}=1$
and $\mathcal{S}^6=1$.
As illustrated in Fig.~\ref{Fig1},
the chiral symmetry has eigenvalues $\pm e^{\pm i\pi/3}, \pm1$.
The topology is characterized by $Z^h\oplus Z^c$ invariants
which are given by
$N_{e^{\pm i\pi/3}} -N_{-e^{\pm i\pi/3}} $ and $N_1-N_{-1}$, respectively.
Similar to the $C_{4z}\mathcal{T}$ case, the BdG Hamiltonian can be block diagonalized as
$H= H_{e^{i2\pi/3}} \oplus H_{e^{-i2\pi/3}} \oplus H_{1} $ according to the eigenvalues $e^{\pm i2\pi/3}, 1$
of $C_{3z}=(C_{6z}\mathcal{T})^2$.
The $H_{e^{i2\pi/3}}$ and $H_{e^{-i2\pi/3}}$ sectors
both belong to the AIII class,
forming a $Z^h$ classification together,
whereas the $H_1$ sector itself forms a $Z^c$ classification (i.e., BDI class) with $\mathcal{P}$
and an effective $\mathcal{T}_{\text{eff}} = (C_{6z}\mathcal{T})^3$ .
%whose square equal to 1.
Therefore, the
topology of the whole BdG Hamiltonian is classified by $Z^h\oplus Z^c$,
which are given by the winding numbers of
 $H_{e^{i2\pi/3}}$  and $H_1$ sectors, respectively.
%Interestingly, the two winding numbers represent helical and chiral end states respectively,
%corresponding to the complex and real eigenvalues of chiral symmetry.
As a consequence, in a 1D superconducting wire with the LMS $C_{6z}\mathcal{T}$, the MZMs can be either chiral or helical,
and the helical and chiral MZMs even can coexist.
Such novel TSC phase may trigger further interests in the manipulation of such helical-chiral coexisting MZMs~\cite{Tanaka2009,Alicea2011,Feng2018}.

\begin{figure}
  \centering
  \includegraphics[width=1\columnwidth]{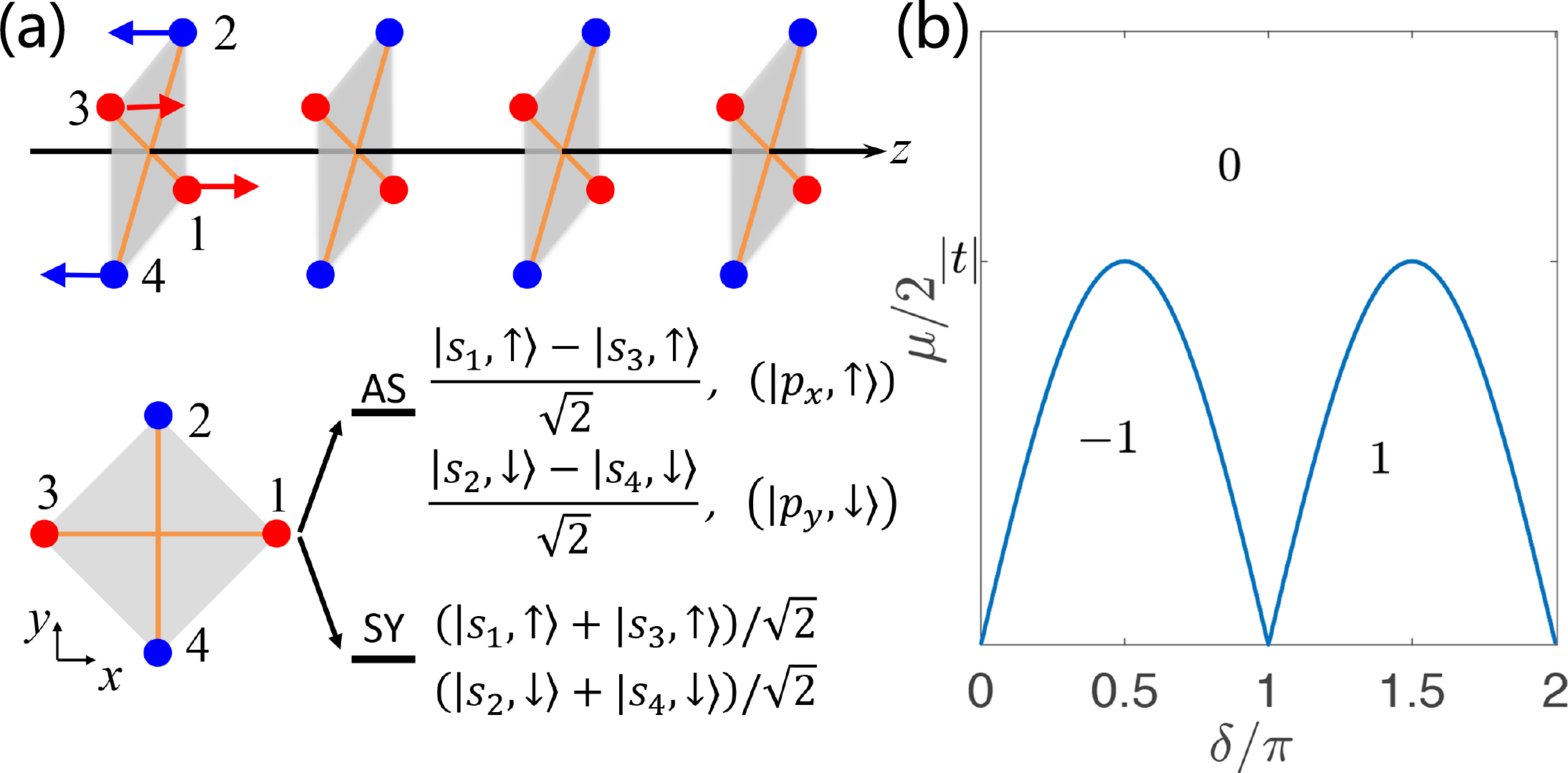}\\
  \caption{(a) A $C_{4z}\mathcal{T}$-preserving superconducting wire aligned along $z$ direction, in which the red and blue dots denote the spin up ($+z$) and spin down ($-z$) polarized $s$-orbitals, respectively. As denoted by the orange bonds, we only introduce the intra-cell coupling between the same spin orbitals, which split the four states into one symmetric (SY) manifold and one antisymmetric (AS) manifold. Both of them are double-degenerate. For simplicity, only the AS manifold is considered in our tight-binding model Eq.(\ref{model}). (b) The topological phase diagram of Eq.(\ref{C4T}) as the function of $\mu$ and $\delta$, in which $0, \pm1$ are the winding numbers, $\mu$ is the chemical potential, and $\delta =\pi/2 + \phi -\alpha$ is the phase difference between the coefficient of $\tau_y$ and $\tau_z$.}\label{Fig2}
\end{figure}

\textit{Model realizations.}---
To illustrate the TSC phase with the LMS $C_{4z}\mathcal{T}$, we construct a 1D anti-ferromagnetic chain along $z$-direction as shown in Fig.~\ref{Fig2}(a), where each unit cell contains four subsites and each subsite is occupied by one spin polarized $s$ orbital. For simplicity, we only consider the intra-cell coupling between the same spin states, which will split the four orbitals into two double-degenerate manifolds as illustrated in Fig.~\ref{Fig2}(a). Taking the anti-symmetric manifold as bases, which equal to the $|p_x,\uparrow\rangle$ and $|p_y,\downarrow\rangle$ states, the effective tight-binding model up to the nearest-neighbor hopping can be written as
\begin{equation}\label{model}
\begin{split}
H_{\text{TB}}^{eff} &= \sum_l t c^{\dag}_{l+1,p_x,\uparrow} c^{}_{l,p_x,\uparrow}
+ t^\ast c^\dag_{l+1,p_y,\downarrow} c^{}_{l,p_y,\downarrow} + h.c.\\
              &+ \mu \sum_{l,\sigma} c^\dag_{l\sigma} c^{}_{l\sigma},
\end{split}
\end{equation}
where $t = |t| e^{i\alpha}$ is the complex hopping,
$\mu$ is the chemical potential and
$\sigma$ acts on the orbital degree of freedom of
$|p_x,\uparrow\rangle$ and $|p_y,\downarrow\rangle$ states.
The $C_{4z}\mathcal{T}$
is given by $ e^{i\pi/4\sigma_z}\sigma_y\mathcal{K}$.
Notice that the hopping terms between opposite spins are prohibited by the $C_{2z}$ symmetry.
The $s$-wave pairing Hamiltonian takes the form
\begin{align}
H_{\Delta} = \sum_l \Delta c^{\dag}_{l+1,p_x,\uparrow} c^{\dag}_{l,p_y,\downarrow}
+ \Delta^\ast c^\dag_{l+1,p_y,\downarrow} c^{\dag}_{l,p_x,\uparrow} + h.c.,
\end{align}
with $\Delta = |\Delta| e^{i\phi}$.
The pairing terms between the same spin are also prohibited by the $C_{2z}$ symmetry.

In the Nambu basis
$(c_{k,p_x,\uparrow}, c^\dag_{-k,p_y,\downarrow},
c_{k,p_y,\downarrow}, c^\dag_{-k,p_x,\uparrow} )^T$,
$\mathcal{P}$ and $\mathcal{T}^\prime$
are given by $\mathcal{P} = \sigma_x \otimes \tau_x \mathcal{K}$
and $\mathcal{T}^\prime = e^{i\pi/4 \sigma_z} \sigma_y \otimes I \mathcal{K}$, respectively,
which give
$\mathcal{S} = e^{-i\pi/4 \sigma_z} \otimes \tau_x $.
The BdG Hamiltonian
anticommutes with $\mathcal{S}$ and takes a block-diagonal form as
\begin{align} \label{eq:eq12}
H_{\text{BdG}}^{C_{4z}\mathcal{T}}  (k)
=
\left(
\begin{array}{cc}
H_{i}(k) &  \\
 & H_{-i}(k) \\
\end{array}
\right),
\end{align}
with
\begin{equation}\label{C4T}
\begin{split}
H_{i}(k) & = |t| \cos(k+\alpha) \tau_z  - |\Delta| \sin(k+\phi) \tau_y + \frac{\mu}{2} \tau_z ,\\
H_{-i}(k) & = |t| \cos(k-\alpha) \tau_z  - |\Delta| \sin(k-\phi) \tau_y + \frac{\mu}{2} \tau_z.
\end{split}
\end{equation}
Then the spectrum is given by
\begin{align}
E(k) = \pm\sqrt{ \left [|t| \cos(k\pm\alpha) +\frac{\mu}{2}\right ]^2  + |\Delta|^2 \sin^2(k\pm\phi) }.
\end{align}
Notice that the two blocks in Eq.~(\ref{eq:eq12})
are Kramers pairs related by the $C_{4z}\mathcal{T}$ symmetry and have the same winding number.
A straightforward way to determine the topology is
to calculate the winding number $v$ by using Eq.~(\ref{eq:eq5}) for the up or lower blocks.
Here we provide a much simpler way to obtain $v$
by analogizing the coefficients of the block Hamiltonians
with elliptically polarized lights
\footnote{The equation describing an elliptically polarized light propagating along $z$-direction is given by $E_x = A_x \cos(kz - \omega t), E_y = A_y \cos(kz - \omega t + \delta)$}.
Taking $H_{i}(k)$ as an example,
the coefficients of the Pauli matrices are
$h_z - \frac{\mu}{2} = |t|\cos (k+\alpha)$ and
$h_y = |\Delta| \sin(k+\phi )$,
which can be regarded as an
elliptically polarized light with amplitude
components $A_x = |t|$, $A_y = |\Delta|$ and
a phase difference  $\delta = \pi/2 + \phi -\alpha$.
When $|t| |\sin \delta| > \frac{\mu}{2}$ ($< \frac{\mu}{2}$),
the
parameter curve of $h_y(k)$ and $h_z(k)$
will (not) wind around the zero point $h_z = h_y = 0$
(We assume $\mu > 0$ for simplicity), and
the superconducting wire is in a topological nontrival (trivial) phase.
Furthermore, when $\delta \in (0,\pi)$ [$\delta \in(-\pi,0)$],
we have a left-handed (right-handed)
parameter curve,
and the topological phase is
characterized by winding number $+1~(-1)$, correspondingly.
The phase diagram in the $\delta - \mu$ parameter space is plotted in Fig.~\ref{Fig2}(b).
We point out that when next-nearest neighbor hopping
and pairing are considered, the competition with nearest neighbor hopping and pairing gives rise to the opportunity for TSC phase with higher winding numbers.

\begin{figure}
  \centering
  \includegraphics[width=1\columnwidth]{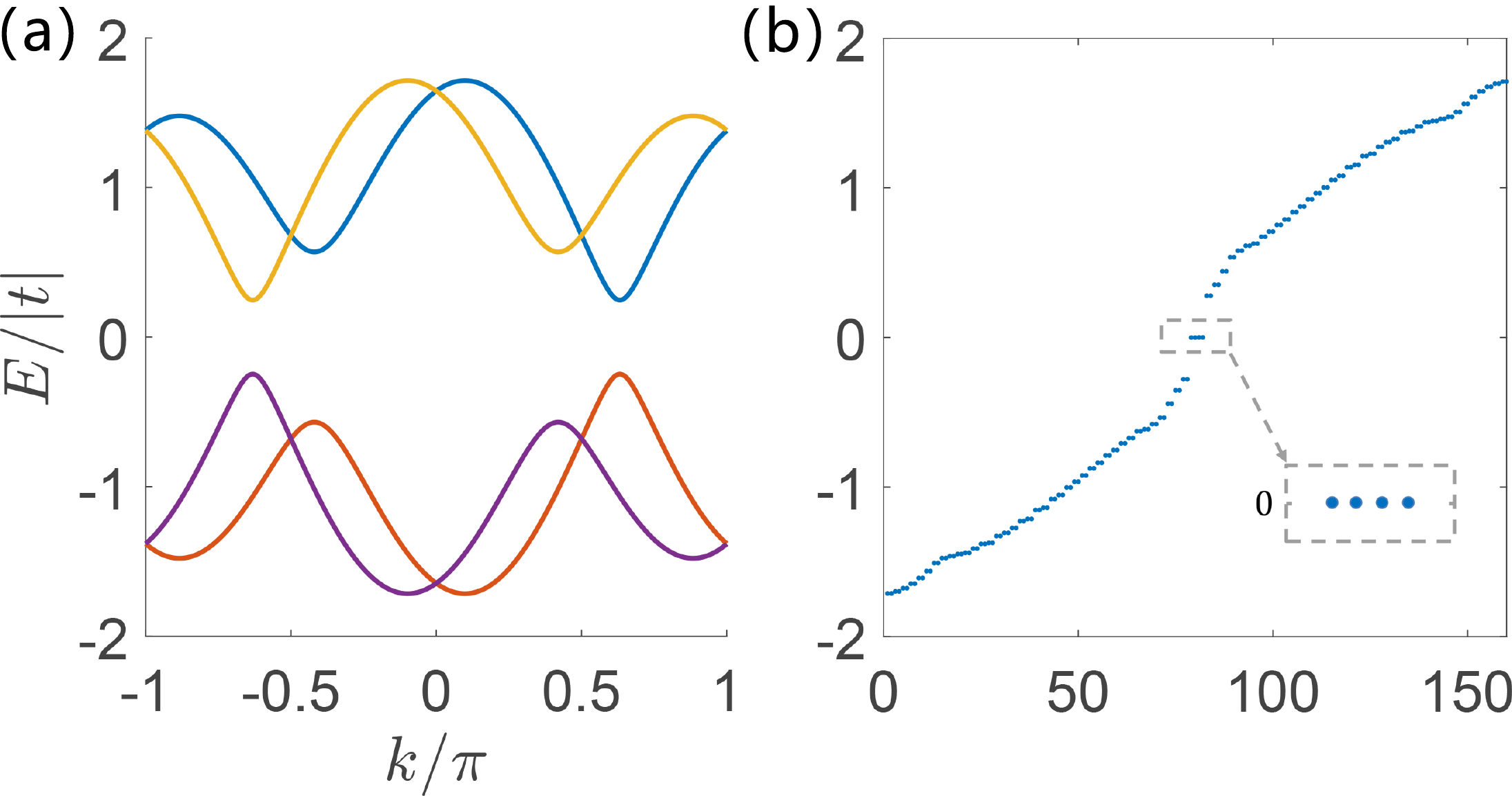}\\
  \caption{The bulk spectrum and MZMs in $C_{4z}\mathcal{T}$-preserving TSC model. (a) is the gapped bulk spectrum of the $C_{4z}\mathcal{T}$-preserving TSC phase with $t=1, \Delta=1.3e^{i\pi/3}, \mu=0.2$, and (b) is the corresponding spectrum of (a) with an open boundary on both sides, in which four MZMs appear at zero energy.}\label{Fig3}
\end{figure}

In the nontrivial TSC phase, the open quantum wire traps an integer pairs of MZMs at its ends.
By using $t=1, \Delta=1.3e^{i\pi/3}, \mu=0.2$, we observe two pairs of MZMs in total on the open wire spectrum, as shown in Fig.~\ref{Fig3}(b),
which is in contrast with the gapped bulk spectrum in Fig.~\ref{Fig3}(a).
These MZMs can also be solved from
the continuous low-energy model~\cite{Bernevig2013}.
Here, we only need to consider $H_{i}(k)$
since the other block in Eq.~(\ref{eq:eq12}),
as well as its zero energy solution,
can be obtained by a  $C_{4z}\mathcal{T}$ transformation.
By assuming the wire is placed on the $z >0$ side,
the low-energy massive Dirac Hamiltonian close to $k = \pi/2$ is given by
\begin{align}
H_{i}
=
\left(
\begin{array}{cc}
 -i|t|\partial_z + \mu/2  &  |\Delta| (-i \sin \delta  + \cos \delta \partial_z ) \\
|\Delta| ( i\sin \delta - \cos \delta \partial_z)  &  i |t| \partial_z - \mu/2 \\
\end{array}
\right).
\end{align}
Its zero energy solution $\Psi_1$
and the $C_{4z}\mathcal{T}$-related partner $\Psi_2 = C_{4z}\mathcal{T}\Psi_1$ are given by
\begin{align}
\Psi_1
=
\left(
\begin{array}{cc}
1 \\
1 \\
0 \\
0 \\
\end{array}
\right) e^{\int dz \frac{i|\Delta| \sin \delta -\mu/2 }{ |\Delta| \cos \delta - i |t| }} ,
\Psi_2
=
\left(
\begin{array}{cc}
0 \\
0 \\
1 \\
1 \\
\end{array}
\right) e^{\int dz \frac{ - i|\Delta| \sin \delta -\mu/2}{ |\Delta| \cos \delta + i |t| }}  .
\end{align}
These two states are the eigenstates of the chiral symmetry $\mathcal{S}$
with eigenvalues $e^{\pm i\pi/4}$, respectively.
Therefore, they are immune to perturbations preserving  $\mathcal{S}$ (and $C_{4z}\mathcal{T}$).
Their combinations give the $\mathcal{S}$ protected MZMs as $\gamma_1 = \Psi_1 + \Psi_2$ and $\gamma_2 = i(\Psi_1 - \Psi_2)$.

The $C_{4z}\mathcal{T}$-preserving BdG Hamiltonian can be easily generalized to a $C_{6z}\mathcal{T}$ invariant quantum wire.
For this purpose,
we assume
the chiral symmetry is expressed as
$\mathcal{S}=(e^{-i\pi/3\tau_z} \otimes \tau_x) \oplus \tau_x$.
%Notice that, for clarity, we used the matrix operation $\otimes$ and $\oplus$ in the expressions.
The BdG Hamiltonian then can be written into three blocks
$H_{\text{BdG}}^{C_{6z}\mathcal{T}} =   H_{e^{i2\pi/3}}(k) \oplus H_{e^{-i2\pi/3}}(k) \oplus H_{1}(k) $
%\begin{equation}\label{eq:eq15}
%    H_{\text{BdG}}^{C_{6z}\mathcal{T}} = \left(
%                         \begin{array}{ccc}
%                           H_{\pm e^{i\pi/3}}(k) & 0 & 0 \\
%                           0 & H_{\pm e^{-i\pi/3}}(k) & 0 \\
%                           0 & 0 & H_{\pm 1} \\
%                         \end{array}
%                       \right),
%\end{equation}
with
\begin{align}
\label{eq:eq16}
\begin{split}
  H_{e^{i2\pi/3}}(k) & =  |t| \cos(k+\alpha)\tau_z - |\Delta| \sin(k+\phi) \tau_y  + \frac{\mu}{2} \tau_z, \\
  H_{e^{-i2\pi/3}}(k) & =  |t| \cos(k-\alpha)\tau_z - |\Delta| \sin(k-\phi) \tau_y  + \frac{\mu}{2} \tau_z, \\
  H_{1}(k) & =  |t'|\cos(k)\tau_z -|\Delta'|\sin(k)\tau_y  + \frac{\mu}{2}\tau_z. \\
\end{split}
\end{align}
The first two blocks are Kramers pairs
and take the same form as in Eq.~(\ref{C4T}),
while the last block is transformed to itself under $C_{6z}\mathcal{T}$ or
$\mathcal{P}$.
For this BdG Hamiltonian,
%the anti-unitary symmetries are represented as
%$C_{6z}\mathcal{T} = [(I\otimes e^{i\pi/3\tau_z})\tau_y \oplus I] \mathcal{K}$
%and $\mathcal{P} = [(\tau_x\otimes\tau_x) \oplus \tau_x]\mathcal{K}$.
the topology is characterized by
$Z^h\oplus Z^c$ numbers, which
correspond to
the number of
helical and chiral MZMs respectively.
The topological phase diagram of the helical part Hamiltonian is the same
as in Fig.~\ref{Fig2}(b).
The chiral part
is determined by the winding number of $H_{1}(k)$,
which gives a nontrivial TSC phase when $|t'|>\frac{\mu}{2}$.

\textit{Conclusion.}---
We have classified the TSC phases of quantum wire with LMSs.
In the case of $M_x\mathcal{T}$ or $C_{2z}\mathcal{T}$, an equivalent BDI class TSC can be realized~\cite{Tewari2012,Mizushima2013,Fang2014}.
More importantly, we find two new types of TSC phases in the superconducting wire with $C_{4z}\mathcal{T}$ or $C_{6z}\mathcal{T}$, which are beyond the already known AZ classes and can be characterized by $Z^h$ or $Z^h\oplus Z^c$ topological invariants, respectively.
These results not only enrich the variety of the 1D TSC, but also provide luxuriant building blocks for the construction of new type 2D and 3D TSCs, by following the general method proposed in Ref.~\cite{Song2018}.

\textit{Acknowledgements.}---
The authors thank Chaoxing Liu for valuable discussion.
This work is supported by the Ministry of Science and Technology of China (No.~2018YFA0307000) and the National Natural
Science Foundation of China (No.~11874022).

\bibliography{TSCs}

\end{document}